\documentclass[fleqn,9pt]{article}
\usepackage[utf8]{inputenc}
\usepackage[T1]{fontenc}
\usepackage{times}
\usepackage{authblk}
\usepackage{fullpage}

\usepackage{xcolor}
\usepackage{amsmath, mathtools, amsfonts, dsfont}
\DeclareMathOperator{\sign}{sign}
\newcommand{\Ee}{\displaystyle {\mathbb E\,}}
\usepackage{tikz, fp}
\usetikzlibrary{calc,arrows, decorations.pathreplacing,decorations.markings, intersections}
\usepackage{xcolor}
\usepackage{nameref,empheq}
\usepackage{pgfplots}
\usepackage{caption}
\usepackage{tkz-euclide} 
\usetkzobj{all} 
\usepackage{graphicx}
\usetikzlibrary{decorations.pathmorphing}
\usepackage{float}
\tikzset{snake it/.style={decorate, decoration=snake}}
\usepackage[square,numbers]{natbib}

\newcommand{\red}[1]{\textcolor{black}{#1}}

\usetikzlibrary{calc,shapes, arrows, arrows.meta, fixedpointarithmetic, shapes.arrows, intersections}

\tikzset{cross/.style={cross out, draw=black, minimum size=2*(#1-\pgflinewidth), inner sep=0pt, outer sep=0pt},
cross/.default={1pt}}

\tikzset{
  on each segment/.style={
    decorate,
    decoration={
      show path construction,
      moveto code={},
      lineto code={
        \path [#1]
        (\tikzinputsegmentfirst) -- (\tikzinputsegmentlast);
      },
      curveto code={
        \path [#1] (\tikzinputsegmentfirst)
        .. controls
        (\tikzinputsegmentsupporta) and (\tikzinputsegmentsupportb)
        ..
        (\tikzinputsegmentlast);
      },
      closepath code={
        \path [#1]
        (\tikzinputsegmentfirst) -- (\tikzinputsegmentlast);
      },
    },
  },
  mid arrow/.style={postaction={decorate,decoration={
        markings,
        mark=at position .5 with {\arrow[#1]{stealth}}
      }}},
}

\makeatletter
\def\calcLength(#1,#2)#3{%
\pgfpointdiff{\pgfpointanchor{#1}{center}}%
             {\pgfpointanchor{#2}{center}}%
\pgf@xa=\pgf@x%
\pgf@ya=\pgf@y%
\FPeval\@temp@a{\pgfmath@tonumber{\pgf@xa}}%
\FPeval\@temp@b{\pgfmath@tonumber{\pgf@ya}}%
\FPeval\@temp@sum{(\@temp@a*\@temp@a+\@temp@b*\@temp@b)}%
\FProot{\FPMathLen}{\@temp@sum}{2}%
\FPround\FPMathLen\FPMathLen5\relax
\global\expandafter\edef\csname #3\endcsname{\FPMathLen}
}
\makeatother

\title{A stochastic approach to colloidal particle collision/agglomeration}

\author[1]{Radu Maftei\thanks{radu.maftei.fr}}
\author[1]{Mireille Bossy\thanks{mireille.bossy@inria.fr}}
\author[3]{Jean-Pierre Minier\thanks{jean-pierre.minier@edf.fr}}
\author[4]{Christophe Profeta\thanks{christophe.profeta@univ-evry.fr}}

\affil[1]{TOSCA Laboratory, INRIA Sophia Antipolis -- M\'editerran\'ee, France}
\affil[3]{EDF R\& D, MEFF, 6, quai Watier, 78400 Chatou, France}
\affil[4]{Laboratoire d'Analyse et Probabilit\'{e}, Universit\'{e} d'Evry-Val d'Essonne, France}

\date{April, 2016}

\begin{document}

\maketitle
\begin{abstract}
Colloidal particles that experience perfectly elastic collisions can be modeled using Langevin processes with specular reflection conditions. The article presents a discretization scheme and offers a conjecture for the rate of convergence of the bias produced. Numerically, these conjectures are confirmed for the specular reflection scheme but also for the absorption scheme, which models perfect agglomeration. 	
\end{abstract}


\medskip

\noindent\rule{\textwidth}{.1pt}
\begin{center} 
\textbf{This paper is accepted in  ICMF-2016 \\
9th International Conference on Multiphase Flow, \\
May 22nd-27th 2016, Firenze, Italy.}
\end{center}

\noindent\rule{\textwidth}{.1pt}
\vspace{2mm}

\section{Introduction}
This paper focuses on modeling and simulating colloidal particle collisions where a stochastic process is driving particle motion. Several approaches have been developed to model such collisions, of which we mention two complementary views:

\begin{itemize}
\item one based on the collision kernel modelling, defined formally as the collision rate divided by the concentration of particles. When the driving stochastic process is a Brownian motion and particle collision results in perfect agglomeration, a practical expression of the collision kernel was already proposed in Ref.\cite{Sm17}. Several extensions are presented in Ref.\cite{Fr77Smoke}. However, it is not straightforward to generalize this expression to more complex situations

\item one based on direct particle tracking simulation \red{where the position-velocity couple is determined for every particle and the particle interactions are explicitly calculated}. More details on these approaches can be found in Ref. \cite{HMP14}
\end{itemize}

The second type of approach is developed in this paper. We consider the case of kinetic particles that follow a Langevin model such as in \cite{MoMi11}:
\begin{equation}\label{eq:Langevin_model}
\begin{dcases}
dx_t = u_t\,dt \\
du_t = \frac{\mathcal{U}_t(x_s) - u_t}{\tau_p}\,dt + B \,dW_t
\end{dcases}
\end{equation}

\noindent where $\tau_p$ is a relaxation time and $(\mathcal{U}_t)_{t \geq 0}$ is the fluid velocity seen by the colloidal particle - it is assumed that the particle has no influence on the fluid velocity field. In the asymptotic $\tau_p$ goes to zero the position process converges towards a Brownian motion, as it is recalled in \nameref{sec:Appendix} in the case of a homogeneous fluid flow.

In the framework of such Langevin models, the analysis of the collision kernel is further developed by considering the case when collisions are followed by specular reflection. Such formulas for the estimation of the collision kernel, based on particle tracking, will later be used in more complex situations such as turbulent flows, where closed-form expressions for the kernel do not exist. 

Models that include a specular type condition have already been introduced in Ref. \cite{DrPo97} which uses such a condition to impose a wall boundary condition on fluid particles in the logarithmic layer. Reference \cite{MiPo99} also uses a specular reflection condition in an alternative approach, when looking for the PDF model equivalent of wall functions.

This paper analyses a simulation scheme for the Langevin model with specular reflection that is similar to the one proposed in Ref. \cite{BeBo10} which used the stochastic Lagrangian approach for fluid particles, and a confirmation of the convergence order proposed by some of the authors in Ref. \cite{BJM16}.

The numerical error of the scheme is considered in a weak sense, meaning that only approximations of any statistic on the particles in position and velocity are considered. We highlight the fact that the precision obtained on these statistics does not rely on the strong approximation of the trajectories as will be discussed in Section~\ref{sec:Discrete}.

The structure of the paper is the following: section~\ref{sec:Model} presents the different stochastic models. Section~\ref{sec:Discrete} presents their discretization and a simulation algorithm. It also offers a conjecture for the rate of convergence in terms of the discretization time step. Section~\ref{sec:Numeric} contains the numeric results that confirm the conjecture and section~\ref{sec:Conclusion} the conclusion.

\section{Model and equations}\label{sec:Model}

 We consider a general model for $N$ mono-dispersed spherical kinetic particles of diameter $\delta$ evolving in a $d$ dimensional domain. For $1 \leq i\leq N$, we denote $z_i=(x_i, u_i)$ the phase space component of particle $i$, with $x_i$ being the center of the particle and $Z_N \coloneqq \{z_1,\ldots, z_N \}$. The phase space of the system is:

\begin{equation}\label{eq:Phase_space_N}
\mathcal{P}^N_\delta \coloneqq \left\lbrace Z_N \in \mathbb{R}^{2dN} ~;~  1\leq i\neq j\leq N,  |x_i - x_j| > \delta \right\rbrace.
\end{equation}  

\noindent We also denote the pre and post collisions borders seen by particle $i$ undergoing a collision with particle $j$:

\begin{align}\label{eq:Phase_border_N}
\partial \mathcal{P}_\delta^{N\pm}(i, j) \coloneqq &\left\lbrace Z_N \in \mathbb{R}^{2dN} ~;~ |x_i - x_j| = \delta , \right.\nonumber\\
								&\quad \pm (u_i-u_j)\cdot(x_i-x_j) > 0, \text{ and } \\
								&\quad \left.\vphantom{\mathbb{R}^{d}} \forall (k, l)\in \{[1,N ]\}^2 \setminus \{(i, j)\}, |x_k-x_l|>\delta \right\rbrace \nonumber
\end{align}

\noindent This form for the collision borders has been chosen such that at any time, a collision can only happen between two and only two particles.

We can now write the equation verified by particle $i$:

	\begin{equation}
	\begin{dcases}
		&\hspace{-1.3em}X^i_t = x^i_0 + \int_0^t U^i_s\,ds\\	
		&\hspace{-1.3em}U^i_t = u^i_0 + \frac{1}{\tau_p}\int_0^t {\color{black}(\mathcal{U}_s(X^i_s) - U^i_s)\,ds} + \tilde{\sigma} W^i_t - \sum_{j \neq i}K_t(i,j)\hspace{-2cm}\\
		&\hspace{-1.3em}K_t(i,j)= \sum_{0< s\leq t}{\color{black}\mathds{1}_{Z^N_s\in \partial\mathcal{P}_\delta^{N+}(i,j)}} \left({\color{black}(U^i_{s-}- U^j_{s-})}\!\cdot\! n_{{\mathcal{P}_\delta^{N}}}(Z^N_s)\right)n_{\mathcal{P}_\delta^{N}}(Z^N_s)\hspace{-2cm} \\		
	\end{dcases} 
	\end{equation}

\noindent where $n_{\mathcal{P}_\delta^{N}}(Z^N_t)$ is the exterior normal to $\mathcal{P}_\delta^{N}$ considered at point $Z^N_t$, the process $(Z^N_t)_{t \geq 0}$ is defined as $\left( X^1_t, U^1_t, \ldots,X^N_t, U^N_t\right)_{t\geq 0}$ and 

\[ \mathds{1}_A = \begin{dcases} 
      1 & \text{ if } A \text{ is true}\\
      0 & \text{ if } A \text{ is false} \\ 
   \end{dcases}
\]

\noindent $A$ being a random variable. Also for all $i$, $(x^i_0, u^i_0) \in \mathbb{R}^{2d}$ are the initial positions and velocities, $(W^i)_{t\geq 0}$ are independent $\mathbb{R}^d$ Brownian motions and $\mathcal{U}_s(X^i_s)$ is the velocity of the fluid seen by the particle $i$. The term $K$ models the jump in the velocity occurring at the perfectly elastic collision between the particles. 

We now isolate two particles, denoted $1$ and $2$ and focus on their relative position and velocity:

\begin{equation}
\begin{dcases}
		& \hspace{-1.3em}X^1_t - X^2_t = x^1_0 - x^2_0 + \int_0^t \left( U^1_s - U^2_s\right)\,ds\\	
		&\hspace{-1.3em}U^1_t - U^2_t = u^1_0 - u^2_0  -\frac{1}{\tau_p}\int_0^t  \left( U^1_s - U^2_s\right)\,ds +\\
		&\qquad\qquad  + \frac{1}{\tau_p}\int_0^t \left(\mathcal{U}_s(X^1_s) - \mathcal{U}_s(X^2_s)\right)\,ds +\\
		&\qquad\qquad + \tilde{\sigma} \left(W^1_t - W^2_t\right) - (K_t(1,2)-K_t(2,1))\\
\end{dcases}
\end{equation}	
	
	\noindent The difference between the perfectly elastic collision terms, particle~$1$ with respect to $2$ and particle~$2$ with respect to $1$, can be developed to obtain:
	
	\begin{align*}
	K_t(1,2)-K_t(2,1) &= 2 \!\!\!\sum_{0< s\leq t}\!\!\! {\color{black}(U^1_{s-}- U^2_{s-})}\cdot n_{{\mathcal{P}_\delta^{2}}}(Z^2_s)) n_{\mathcal{P}_\delta^{2}}(Z^2_s)\mathds{1}_{Z^2_s\in \partial\mathcal{P}_\delta^{2+}(1,2)}
	\end{align*}
	
	\noindent by using the fact that $ \mathcal{P}_\delta^{2+}(1,2) = \mathcal{P}_\delta^{2+}(2,1)$. This shows that the process $(X^1-X^2, U^1-U^2)$ has a specular reflection condition at the collision border.

The relative distance is denoted as $X_t \coloneqq X^1_t - X^2_t$, the relative velocity as $U_t \coloneqq U^1_t - U^2_t$. Since the Brownians $W^1$ and $W^2$ are independent, we introduce a new Brownian $(W_t)_{t\geq 0}$ such that $W_t \coloneqq \frac{1}{\sqrt{2}}(W^1_t - W^2_t)$. Let $\sigma \coloneqq \sqrt{2}\tilde{\sigma}$. For the purpose of the analysis, we also assume that the fluid velocity can be linearized so that the drift term can be replaced by a generic function $b(t, X^1_t - X^2_t, U^1_t - U^2_t)$ modeling the term:

	\begin{align*}
	\frac{1}{\tau_p} \left(\mathcal{U}_t(X^1_t) - \mathcal{U}_t(X^2_t) \right) -\frac{1}{\tau_p} \left(U^1_t - U^2_t \right)
	\end{align*}
	
	 The phase space for the relative process is:
\begin{equation}\label{eq:Phase_space_wall}
\mathcal{P} \coloneqq \left\lbrace (x, u) \in \mathbb{R}^{2d}  ~;~  |x| > \delta \right\rbrace
\end{equation}  
	
\noindent with pre and post collision borders:
	\begin{align}\label{eq:Phase_border_wall}
\partial \mathcal{P}^{\pm} \coloneqq &\left\lbrace (x, u) \in \mathbb{R}^{2d}~;~|x| = \delta , \pm u\cdot x > 0\right\rbrace
\end{align}
	
	\noindent We note $\mathcal{D}$ the configuration space and $\partial \mathcal{D}$ its border. 

So using these notations the model becomes:
		
	\begin{equation}\label{eq:Langevin_specular}
	\begin{dcases}
		& X_t = x+ \int_0^t U_s\,ds\\	
		& U_t = u + \int_0^t{\color{black} b(s, X_s, U_s)}\,ds + \sigma W_t - K_t\\
		&K_t = \sum_{0< s\leq t} 2(U_{s-}\cdot n_{\mathcal{D}}(X_s))n_{\mathcal{D}}(X_s) \mathds{1}_{X_s \in \partial \mathcal{D}} 
	\end{dcases}
	\end{equation}
	
	\subsection{Known mathematical results about the specular reflection}
	
	In the following, we will consider our system up to a finite time $T$.
	
	In the unidimensional case $d=1$, in Ref. \cite{BossyJabir11} it is shown that when the drift $b$ follows a homogeneity condition with respect to the sign of $x$, meaning $b(x, u) = \sign(x) b(|x|, \sign(x) u)$ then the solution of Eqn. \eqref{eq:Langevin_specular} is exactly given by:

\begin{empheq}[left = \empheqlbrace]{align}
        &\quad X_t = |X^f_t| \label{eq:Exact_Solution_1}\\
		&\quad U_t = \sign(X^f_t) U^f_t \label{eq:Exact_Solution_2}      
\end{empheq}

	 \noindent where $(X^f_t, U^f)_{t\geq 0}$ solve the following free Langevin equation:
	\begin{equation}\label{eq:Langevin_Free}
	\begin{dcases}
		& X^f_t = x + \int_0^t U^f_s\,ds\\	
		& U^f_t = u + \int_0^t{\color{black} b(s, X^f_s, U^f_s)}\,ds + \sigma W_t
	\end{dcases}
	\end{equation}
	free meaning that the position component does not see any collision borders.

	The authors also show that the sum defined in the collision term $K$ in Eqn. \eqref{eq:Langevin_specular} is well posed -meaning the set of collision times is countable- if the process does not start from the initial position $(x_0, u_0)=(0, 0)$.

	%Reference \cite{BossyJabir15} shows that if one considers 
	Consider any statistic depending on the starting state~$(t, x, u)$: $\Gamma (t, x, u) = \Ee \psi(X^{t,x, u}_T, U^{t,x, u}_T)$  where the process $(X^{t,x, u}_T, U^{t,x, u}_T)_{[t, T]}$ follows Eqn \eqref{eq:Langevin_specular} with starting condition at time $t$: $(X^{t,x, u}_t, U^{t,x, u}_t) = (x, u)$.  Then, by Ref. \cite{BossyJabir15},  $\Gamma$ is the weak solution of the following PDE:

%REGULARITY INSIDE THE DOMAIN	
	
	\begin{equation}\label{eq:Main_PDE}
	\begin{dcases}
	&\hspace{-1em}\frac{\partial \Gamma}{\partial t} + u\nabla_x \Gamma + b(t, x, u)\nabla_u \Gamma + \frac{\sigma^2}{2}\Delta_u\Gamma = 0 \text{ on }[0,T]\times \mathcal{P}\hspace{-2cm}\\
	&\hspace{-1em}\Gamma(T, x, u) = \psi(x, u) \text{ on } \mathcal{P}\hspace{-2cm} \\
	&\hspace{-1em}\Gamma(t, x, u) = \Gamma(t, x, u- 2(u\cdot n_{\mathcal{D}(x)}) n_{\mathcal{D}(x)}) \text{ on } [0,T]\times \partial\mathcal{P}^{+}\hspace{-2cm}		
	\end{dcases}
	\end{equation}	

\noindent assuming $\psi$ is a smooth function with compact support.

	One can notice therefore that the solution of the PDE defined from the expectation of the process also follows a specular boundary condition.

	The PDE \eqref{eq:Main_PDE} is used to obtain a reference result for the numerical experiments as explained in Section \ref{sec:Numeric}.

\section{Discrete simulation schemes}\label{sec:Discrete}

The schemes used in the numerical experiments are presented in the section. The main focus is on the specular reflection scheme but an absorption scheme is also be briefly presented. 

\subsection{Specular reflection scheme}

The details for the unidimensional scheme are exposed and the results of the numerical simulations will be given also in this setting. In Subsection~\ref{ssec:MultiD_Specular} the multidimensional framework is then briefly deduced.

In the  unidimensional version of Eqn \eqref{eq:Langevin_specular}, the collision border is like a fixed wall at a given position $\delta$. Without any loss of generality, we will assume in this section that $\delta \equiv 0$. The equation can be written as:

	\begin{equation}\label{eq:Langevin_specular_1D}
	\begin{dcases}
		& X_t = x+ \int_0^t U_s\,ds\\	
		& U_t = u + \int_0^t{\color{black} b(s, X_s, U_s)}\,ds + \sigma W_t - K_t\\
		&K_t = \sum_{0< s\leq t} 2 U_{s-} \mathds{1}_{X_s = 0} 
	\end{dcases}
	\end{equation}

Let $0 = t_0 < t_1 < \ldots < t_n = T$ be a uniform partition of $[0, T]$ of time step $t_{i+1}- t_i \coloneqq \Delta t$. The discretized process $(\bar{X}_t, \bar{U}_t)_{t\in[0, T]}$ is given here in a time continuous formulation but, as shown later on, this process is exactly simulable at the time steps that matter. The construction  is iterative: set $(\bar{X}_0, \bar{U}_0) = (x_0, u_0)$.  Just as the process in \eqref{eq:Exact_Solution_1}-\eqref{eq:Exact_Solution_2} solves Eqn. \eqref{eq:Langevin_specular} for some drift $b$, a process that mimics the position component of the free Langevin model is introduced and denoted as {\color{black}$(\bar{X}^f_t)_{t\in[0, T]}$}. This process starts as $\bar{X}^f_0 = x_0$.  

Therefore, the discretization of the position component of Eqn. \eqref{eq:Langevin_specular_1D}, between two time steps $t_i$ and $t_{i+1}$, is:
	
	\begin{align}\label{eq:Position_discretization}
		\left\lbrace
		\begin{array}{l}
		\text{if } t_i \leq t \leq t_{i+1}\colon\\
		\qquad \bar{X}^f_{t} = \bar{X}_{t_i} + (t- t_i)\bar{U}_{t_i} \\
		\qquad \bar{X}_{t} = |\bar{X}^f_{t}|
		\end{array}
		\right.
	\end{align}
solving exactly Eqn. \eqref{eq:Exact_Solution_1} with a constant velocity process $\bar{U}_{t_i}$. The position process satisfies the constraint of remaining in the domain $\mathcal{D}$ at any time. 

The hitting time of the wall at zero \red{of $(\bar{X}_t)_{t\geq 0}$} is defined as:

\begin{equation}\label{eq:Hitting_time}
\theta_i = \min\left(\max\left( t_i, \left(t_i - \frac{\bar{X}_{t_i}}{\bar{U}_{t_i}} \right)\right), t_{i+1}\right)
\end{equation}

This means that on the time step $[t_i, t_{i+1}]$ the particle moves at constant velocity $\bar{U}_{t_i}$ if it does not hit the reflection border, therefore it is know at time $t_i$ if the particle will hit or not the border. And if the specular reflection border is to be hit, it happens at the instant $t_i - \bar{X}_{t_i}/\bar{U}_{t_i}$. The min-max taken afterwards is simply to make the hitting time be in the time step if it does not hit the border between the times $t_i$ and $t_{i+1}$. As such, the scheme considered has at most one collision per time step.

A collision is detected when $t_i < \theta_i < t_{i+1}$. The inclusion of collisions in the scheme is the reason why continuous processes $(\bar{X}_t, \bar{U}_t, Z_t)$ are presented. $\theta_i$ is known at the beginning of the time step $t_i$ for the interval $[t_i, t_{i+1})$ but it is unknown before. 

Concerning the discretization of the velocity, we set that:
		
	\begin{align}\label{eq:Velocity_discretization}
	\left\lbrace
		\begin{array}{l}
		\text{if }t_i < \theta_i < t_{i+1} \text{ (collision detected) }\colon  \\ 
		\quad  \text{if } t_i\leq  t < \theta_i \\
			%\qquad 		 
		 \qquad \bar{U}_{t} = \bar{U}_{t_i} + b(\bar{X}_{t_i}, \bar{U}_{t_i})(t-t_i) + \sigma Z^-_i(t_i, t)\\ 
		\quad\text{reflection}\colon \\
			\qquad \bar{U}_{\theta_i} = -\bar{U}_{\theta_i^-}\\
		\quad \text{if } \theta_i \leq t < t_{i+1}\colon\\
		  \qquad \bar{U}_{t} = \bar{U}_{\theta_i} + b(\bar{X}_{\theta_i}, \bar{U}_{\theta_i})(t-\theta_i) + \sigma Z^+_i(\theta_i, t)\\
		\text{else (no collision)}\colon\\
		\quad \text{if } t_i \leq t< t_{i+1}\colon\\
		\qquad \bar{U}_{t} = \bar{U}_{t_i} + b(\bar{X}_{t_i}, \bar{U}_{t_i})(t-t_i) + \sigma Z_i(t_i, t)\\
		\end{array} 
		\right.
	\end{align}

\noindent where $Z^-_i(t_i, t)$, $Z^+_i(\theta_i, t)$ and $Z_i(t_i, t)$ are independent random processes that have the same Gaussian distribution as $W_t - W_{t_i}$, $W_t - W_{\theta_i}$ and respectively $W_t - W_{t_i}$.

While the scheme is described in continuous time, the only points of the scheme that matter are $(\bar{X}_{t_i}, \bar{U}_{t_i}, \bar{U}_{\theta_i})_{i\in \{0,\ldots, n\}}$, so for Eqn.~\eqref{eq:Velocity_discretization} only $Z^-_i(t_i, \theta_i)$, $Z^+_i(\theta_i, t_{i+1})$ and $Z_i(t_i, t_{i+1})$ are needed. We then have that  $Z^-_i(t_i, \theta_i) = \sqrt{\theta_i-t_i} Z^-_i$,  $Z^+_i(\theta_i, t_{i+1}) = \sqrt{t_{i+1}-\theta_i} Z^+_i$ and $Z_i(t_i, t_{i+1}) = \sqrt{t_{i+1} - t_i}Z_i$, where $Z^-_i$, $Z^+_i$ and $Z_i$ are independent Gaussian random variables with mean 0 and variance 1.

One can see that on a period $[t_i, t_{i+1}]$, there are two velocities attached to the position $\bar{X}_t$. The first, appearing in Eqn. \eqref{eq:Position_discretization}, is used to determine the position of the particle at each instant in the time step. The second which appears in Eqn. \eqref{eq:Velocity_discretization} is used to calculate the starting velocity for the time step $[t_{i+1}, t_{i+2}]$ but has no influence on the position in the current interval. \textit{As such, $U_{\theta_i^-}$ is not the velocity at which the particle hits the border, and it is in this sense that the approximation given by this scheme is not a trajectorial one.} 

In order to illustrate this, we plot what the scheme does on the time step $[t_i, t_{i+1}]$. We assume that $b \equiv 0$ so as to highlight the influence of the Brownian noise and of the specular reflection. The upper axis plots the position across time and the velocity at which this position changes. The lower axis plots the velocity component of Eqn. \eqref{eq:Velocity_discretization} at time $t_i$, $\theta_i$ and $t_{i+1}$. The velocity $\bar{U}_{t_i}$ in the lower part is equal to velocity in the upper part. In both schemes $Z^-\coloneqq Z^-_i(t_i, \theta_i)$ and $Z^-\coloneqq Z^+_i(\theta_i, t_{i+1})$.

\begin{figure}[H] 
\centering
\begin{tikzpicture}

\def\disXU{3}
\coordinate (ti) at (1, 0);
\coordinate (theta) at (4, 3);
\coordinate (tf) at (6, 0);

%ti
\node at (ti) [below=5pt]{$t_i$};
\draw[thick] ($(ti)+(0,4pt)$) -- ($(ti)-(0,4pt)$);
\draw[thick] ($(ti)+(0, \disXU)+(0,4pt)$) -- ($(ti)+(0, \disXU)-(0,4pt)$);
\draw[dashed](ti)--++(0, \disXU);

%tf
\node at (tf) [below=5pt]{$t_{i+1}$};
\draw[thick] ($(tf)+(0,4pt)$) -- ($(tf)-(0,4pt)$);
\draw[thick] ($(tf)+(0, \disXU)+(0,4pt)$) -- ($(tf)+(0, \disXU)-(0,4pt)$);
\draw[dashed](tf)--++(0, \disXU);

%theta i
\node at (theta) [above=5pt]{$\theta_i$};
\draw[thick] ($(theta)+(0,4pt)$) -- ($(theta)-(0,4pt)$);
\draw[thick] ($(theta)-(0, \disXU)+(0,4pt)$) -- ($(theta)-(0, \disXU)-(0,4pt)$);
\draw[dashed](theta)--(theta |-{{(0,0)}});

\draw[-latex, thick](0,0)node[left]{$0$} -- ++(7, 0)node[right]{$t$};
\draw[-latex, thick](0, -2) -- ++(0, 4.5)node[right]{$\bar{U}$};

\draw[-latex, thick](0,\disXU) -- ++(7, 0)node[right]{$t$};
\draw[-latex, thick](0,\disXU) -- ++(0, 2)node[right]{$\bar{X}$}; 

%X
\coordinate (Xi) at ($(ti)+(0, 4.5)$);
\coordinate (Xf) at (6, 4);

\node at (Xi) [above=0.2cm]{${\bar{X}_{t_i}}$};
\node at (Xf) [above=0.2cm]{${\bar{X}_{t_{i+1}}}$};
\draw (Xi) node[cross=4pt, very thick]{};
\draw (Xf) node[cross=4pt, very thick]{};

\begin{scope}[very thick,decoration={
    markings,
    mark=at position 0.5 with {\arrow[black]{>}}}
    ] 
    \draw [very thick, postaction={decorate}] (Xi) -- node[above, yshift=0.2cm]{${\bar{U}_{t_i}}$}(4,\disXU);
    \draw [very thick, postaction={decorate}] (4,\disXU) --  node[above, yshift=0.2cm]{${-\bar{U}_{t_i}}$}(Xf);
\end{scope} 

\coordinate (Ui) at ($(ti)+(0, -1.5)$);
\coordinate (Utheta) at ($(theta)-(0, 4)$);
\coordinate (UthetaR) at ($(Utheta) + (0, 2)$);
\coordinate (Uf) at ($(tf) + (0, 2)$);

\draw (Ui) node[cross=4pt, very thick]{};
\draw (Utheta) node[cross=4pt, very thick]{};
\draw (UthetaR) node[cross=4pt, very thick]{};
\draw (Uf) node[cross=4pt, very thick]{};

\node at (Ui) [right=0.2cm]{${\bar{U}_{t_i}} $};
\node at (Utheta) [right=0.2cm]{${\bar{U}_{\theta^-_i}}$};
\node at (UthetaR) [right=0.2cm]{${\bar{U}_{\theta_i}}$};
\node at (Uf) [right=0.2cm]{${\bar{U}_{t_{i+1}}}$};

\draw[very thin, loosely dashed](Ui)--(Ui -|{{(0,0)}});
\draw[very thin, loosely dashed](Utheta)--(Utheta -|{{(0,0)}});
\draw[very thin, loosely dashed](UthetaR)--(UthetaR -|{{(0,0)}});
\draw[very thin, loosely dashed](Uf)--(Uf -|{{(0,0)}});

 \draw [thick, decorate,decoration={brace,amplitude=2pt, raise=5pt}](Ui -|{{(0,0)}})-- (Utheta -|{{(0,0)}}) node[black,midway,xshift=-0.6cm] {$Z^-$};
 \draw [thick, decorate,decoration={brace,amplitude=2pt, raise=5pt}](UthetaR -|{{(0,0)}})-- (Uf -|{{(0,0)}}) node[black,midway,xshift=-0.6cm] {$Z^+$};
 
 \draw[->, thick, dotted, shorten >=0.2cm, shorten <=0.2cm] (Utheta)[bend left] to (UthetaR) ;
\end{tikzpicture}
\caption{Evolution of the scheme $(\bar{X}_t, \bar{U}_t)$ assuming $b\equiv 0$}	
\label{fig:Specular_Scheme}
\end{figure}
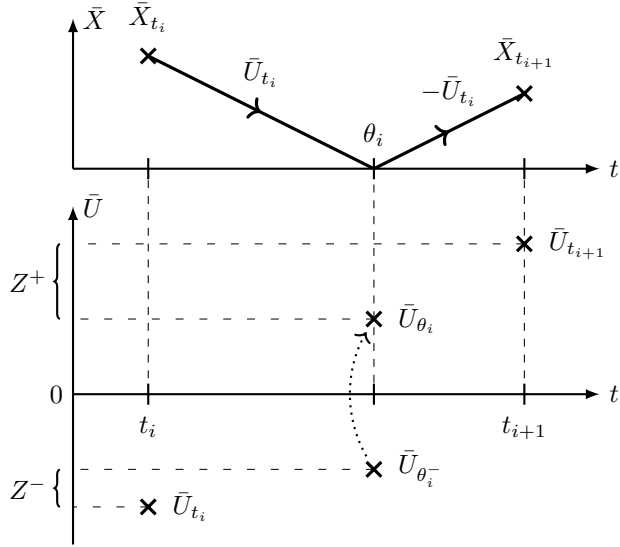

One can notice in Fig. \ref{fig:Specular_Scheme} that the instant $\theta_i$ is defined by the position process. At this instant, one determines how much the velocity has diffused between $[t_i, \theta_i)$ by calculating $\bar{U}_{\theta_i^-}$ which is flipped to simulate the specular reflection. Afterwards the velocity process diffuses until the end of the time step. 

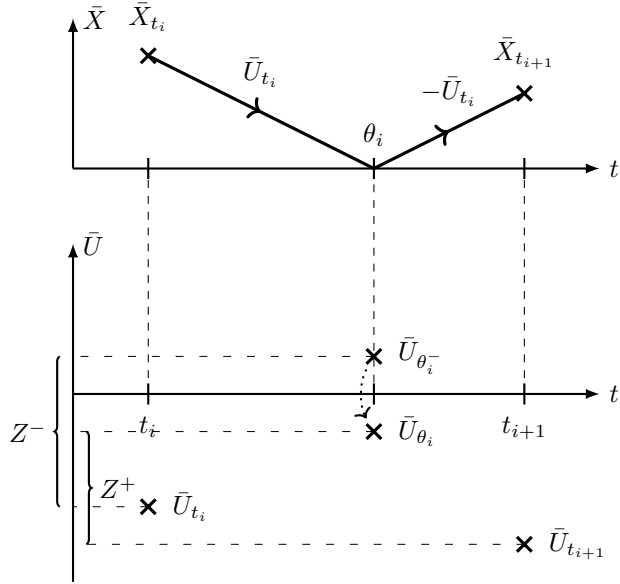
\begin{figure}[H]
\centering
\begin{tikzpicture}
%\draw[step=1cm,gray,very thin, dashed] (0,0) grid (8,4);
\def\disXU{3}
\coordinate (ti) at (1, 0);
\coordinate (theta) at (4, 3);
\coordinate (tf) at (6, 0);

%ti
\node at (ti) [below=5pt]{$t_i$};
\draw[thick] ($(ti)+(0,4pt)$) -- ($(ti)-(0,4pt)$);
\draw[thick] ($(ti)+(0, \disXU)+(0,4pt)$) -- ($(ti)+(0, \disXU)-(0,4pt)$);
\draw[dashed](ti)--++(0, \disXU);

%tf
\node at (tf) [below=5pt]{$t_{i+1}$};
\draw[thick] ($(tf)+(0,4pt)$) -- ($(tf)-(0,4pt)$);
\draw[thick] ($(tf)+(0, \disXU)+(0,4pt)$) -- ($(tf)+(0, \disXU)-(0,4pt)$);
\draw[dashed](tf)--++(0, \disXU);

%theta i
\node at (theta) [above=5pt]{$\theta_i$};
\draw[thick] ($(theta)+(0,4pt)$) -- ($(theta)-(0,4pt)$);
\draw[thick] ($(theta)-(0, \disXU)+(0,4pt)$) -- ($(theta)-(0, \disXU)-(0,4pt)$);
\draw[dashed](theta)--(theta |-{{(0,0)}});

\draw[-latex, thick](0,0) -- ++(7, 0)node[right]{$t$};
\draw[-latex, thick](0, -2.5) -- ++(0, 4.5)node[right]{$\bar{U}$};

\draw[-latex, thick](0,\disXU) -- ++(7, 0)node[right]{$t$};
\draw[-latex, thick](0,\disXU) -- ++(0, 2)node[right]{$\bar{X}$}; 

%X
\coordinate (Xi) at ($(ti)+(0, 4.5)$);
\coordinate (Xf) at (6, 4);

\node at (Xi) [above=0.2cm]{${\bar{X}_{t_i}}$};
\node at (Xf) [above=0.2cm]{${\bar{X}_{t_{i+1}}}$};
\draw (Xi) node[cross=4pt, very thick]{};
\draw (Xf) node[cross=4pt, very thick]{};

\begin{scope}[very thick,decoration={
    markings,
    mark=at position 0.5 with {\arrow[black]{>}}}
    ] 
    \draw [very thick, postaction={decorate}] (Xi) -- node[above, yshift=0.2cm]{${\bar{U}_{t_i}}$}(4,\disXU);
    \draw [very thick, postaction={decorate}] (4,\disXU) --  node[above, yshift=0.2cm]{${-\bar{U}_{t_i}}$}(Xf);
\end{scope} 

\coordinate (Ui) at ($(ti)+(0, -1.5)$);
\coordinate (Utheta) at ($(theta)-(0, 2.5)$);
\coordinate (UthetaR) at ($(Utheta) - (0, 1)$);
\coordinate (Uf) at ($(tf) + (0, -2)$);

\draw (Ui) node[cross=4pt, very thick]{};
\draw (Utheta) node[cross=4pt, very thick]{};
\draw (UthetaR) node[cross=4pt, very thick]{};
\draw (Uf) node[cross=4pt, very thick]{};

\node at (Ui) [right=0.2cm]{${\bar{U}_{t_i}}$};
\node at (Utheta) [right=0.2cm]{${\bar{U}_{\theta^-_i}}$};
\node at (UthetaR) [right=0.2cm]{${\bar{U}_{\theta_i}}$};
\node at (Uf) [right=0.2cm]{${\bar{U}_{t_{i+1}}}$};

\draw[very thin, loosely dashed](Ui)--(Ui -|{{(0,0)}});
\draw[very thin, loosely dashed](Utheta)--(Utheta -|{{(0,0)}});
\draw[very thin, loosely dashed](UthetaR)--(UthetaR -|{{(0,0)}});
\draw[very thin, loosely dashed](Uf)--(Uf -|{{(0,0)}});

 \draw [thick, decorate,decoration={brace,amplitude=2pt, raise=5pt}](Ui -|{{(0,0)}})-- (Utheta -|{{(0,0)}}) node[black,midway,xshift=-0.6cm] {$Z^-$};
 \draw [thick, decorate,decoration={brace,amplitude=2pt, raise=5pt}](UthetaR -|{{(0,0)}})-- (Uf -|{{(0,0)}}) node[black,midway,xshift=0.6cm] {$Z^+$};
 
  \draw[->, thick, dotted, shorten >=0.2cm, shorten <=0.2cm] (Utheta)[bend right] to (UthetaR)  ;
\end{tikzpicture}	
\caption{Evolution of the scheme, flipping positive velocity}
\label{fig:Specular_scheme2}
\end{figure}

Since the velocity is free to diffuse, it is possible that at time $\theta_i$, it is positive. This case is plotted in Fig. \ref{fig:Specular_scheme2}. The velocity processes diffuses up to $\theta_i$ and the result is: $\bar{U}_{\theta^-_i} > 0$. Even in this case, we flip the velocity in a specular reflection way. Again, one can see both in Fig. \ref{fig:Specular_Scheme} and in \ref{fig:Specular_scheme2} that the trajectory of the particle depends only on the position and velocity at the beginning of the time step $(\bar{X}_{t_i}, \bar{U}_{t_i})$. 
\subsection{Weak error of the scheme}

While the full simulation algorithm utilises a Monte Carlo procedure to compute $\Ee f(\bar{X}_T, \bar{U}_T)$ -leading to an additional variance error- in this section we only consider the bias error defined as:

\begin{equation}\label{eq:Error}
\text{Bias}[f](\Delta t) \coloneqq \Ee f(X_T, U_T) - \Ee f(\bar{X}_T, \bar{U}_T) 
\end{equation} 

\noindent for any $f$, a smooth function.

A bias error on the whole trajectory, called strong error, would have meant measuring the difference on the whole time domain $[0, T]$ as:

\begin{equation}\label{eq:Strong_Error}
\text{Bias}_\text{strong}(\Delta t) \coloneqq \Ee \left|X_T - \bar{X}_T  \right| + \Ee \left|U_T - \bar{U}_T  \right| 
\end{equation}

Bias$[f]$ quantifies the approximation error between the distribution of $(X_T, U_T)$ and the distribution of $(\bar{X}_T, \bar{U}_T)$ while $\text{Bias}_\text{strong}$ quantifies the approximation error between the random variables $(X_T, U_T)$ and $(\bar{X}_T, \bar{U}_T)$ constructed using the same trajectories of the Brownian motion $(W_t)_{t\geq 0}$.

Evidently, as soon as $f$ is Lipschitz continuous:

\[
\left|\text{Bias}[f](\Delta t)\right| \leq L_f \text{Bias}_\text{strong}(\Delta t)
\]

\noindent where $L_f$ is the Lipschitz constant of $f$.
 
It is also well know (see Ref. \cite{BGT04} and \cite{GrTa13}) when the diffusion coefficient $\sigma$ is not constant or/and when boundary conditions are imposed on the process, that the rate of convergence of  $\text{Bias}[f]$ is of higher order than the one of $\text{Bias}_\text{strong}$.

While in Ref. \cite{BJM16}, the authors prove that:
\begin{equation}\label{eq:Error_estimate_ref}
\left|\text{Bias}[f](\Delta t)\right| \leq \mathcal{O}(\Delta t)
\end{equation}

\noindent we conjecture that:
\begin{equation}\label{eq:Error_estimate}
\text{Bias}[f](\Delta t) = C \Delta t  + \mathcal{O}\left(\Delta t^2\right)
\end{equation}
\noindent $\Delta t$ being the discretization step of $[0, T]$ and $C$ a constant. 

When expanded it becomes:

\begin{equation}\label{eq:Romberg_1}
\Ee f(X_T, U_T) - \Ee f(\bar{X}_T, \bar{U}_T) = {C}\Delta t + \mathcal{O}\left(\Delta t^2\right)
\end{equation}

\noindent If one takes the same scheme $(\bar{X}_t^{\Delta t/2}, \bar{U}_t^{\Delta t/2})$ but with time increment $\Delta t/2$ then one has:

\begin{equation}\label{eq:Romberg_2}
\Ee f(X_T, U_T) - \Ee f(\bar{X}_T^{\Delta t/2}, \bar{U}_T^{\Delta t/2}) = {C}\frac{\Delta t}{2} + \mathcal{O}\left(\Delta t^2\right)
\end{equation}
Multiplying Eqn.~\eqref{eq:Romberg_2} and subtracting from it Eqn.~\eqref{eq:Romberg_1}, one obtains:

\begin{equation*}
\Ee f(X_T, U_T) - 2\Ee f(\bar{X}_T^{\Delta t/2}, \bar{U}_T^{\Delta t/2}) + \Ee f(\bar{X}_T, \bar{U}_T) = \mathcal{O}\left(\Delta t^2\right)
\end{equation*}

So if the convergence rate is linear then by combining two simulation results with different time increments, one obtains a quadratic convergence rate. This result is the Richardson-Romberg extrapolation which will be used later on.

\subsection{Multidimensional specular reflection scheme}\label{ssec:MultiD_Specular}

The multidimensional setting can be derived from the unidimensional case as follows: at time $t_i$, the free position $\bar{X}^f_{t_{i+1}} =  \bar{X}_{t_i} + \Delta t\bar{U}_{t_i}$ is calculated. If the linear path from $\bar{X}_{t_i}$ to $\bar{X}^f_{t_{i+1}}$ crosses the domain $\mathcal{D}^c~=~\{x~\in~\mathbb{R}^d~;~|x|~\leq~\delta\}$, then a collision is detected. By determining the first intersection of the free process with the $\partial \mathcal{D}$, one obtains the collision point $\bar{X}_{\theta_i}$ (where $\|\bar{X}_{\theta_i}\| = \delta$) and the collision time $\theta_i$. The collision point determines the collision plane, tangential to the border $\partial\mathcal{D}$ at $\bar{X}_{\theta_i}$. 

The position $\bar{X}_{t_{i+1}}$ is determined by effectuating a mirror reflection of $\bar{X}^f_{t_{i+1}}$ with respect to the collision plane as in Eqn.~\eqref{eq:Exact_Solution_1}. 

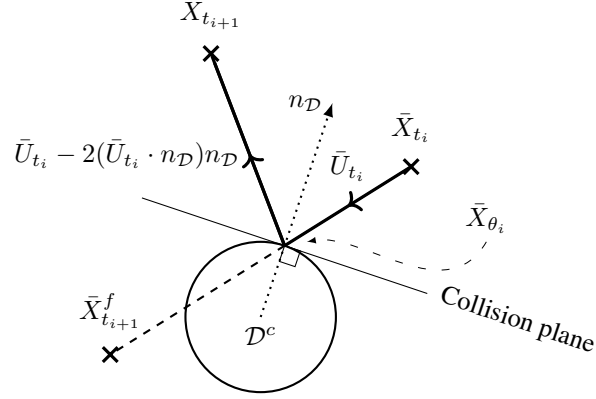
\begin{figure}[H]
\centering
\begin{tikzpicture}[dot/.style={circle,inner sep=1pt,fill,label={#1},name=#1},
 extended line/.style={shorten >=-#1,shorten <=-#1},
 extended line/.default=1cm,
 one end extended/.style={shorten >=-#1},
 one end extended/.default=1cm,
 ]

\coordinate (O) at (3, 1);
\draw[thick, name path =circ1] (O) circle (1cm) node[below] {$\mathcal{D}^c$};

\coordinate (X0) at (5, 3);
\coordinate (X1) at (1, 0.5);
\node at (X1)[cross=4pt, very thick]{};
\node at (X1)[above=0.2cm]{${\bar{X}^f_{t_{i+1}}}$};
\node at (X0)[cross=4pt, very thick]{};
\node at (X0)[above=0.2cm]{${\bar{X}_{t_{i}}}$};

\path[name path=xfree](X0)--(X1);

\path [name intersections={of=circ1 and xfree,by=Xtheta}];
\draw[thick](X0) -- (Xtheta);

\begin{scope}[very thick,decoration={
    markings,
    mark=at position 0.5 with {\arrow[black]{>}}}
    ] 
    \draw [very thick, postaction={decorate}] (X0) --  node[above, yshift=0.2cm]{${\bar{U}_{t_i}}$}(Xtheta);
\end{scope} 

\draw[thick, dashed](Xtheta)--(X1);

\draw[thin] (Xtheta) -- ($(Xtheta)!2!90:(O)$);
\draw[thin] (Xtheta) -- ($(Xtheta)!2!-90:(O)$);

\coordinate (Xtg) at ($(Xtheta)!2!90:(O)$);
\path (Xtheta) -- (Xtg) node [midway, above, sloped, xshift=2.25cm, yshift=-0.25cm] {Collision plane};
\tkzFindAngle(X1,Xtheta,O)
\tkzGetAngle{angleRefl}

\tkzMarkRightAngle(O,Xtheta,Xtg);

\path let \p1=(X1), \p2=(Xtheta), \n1={veclen(\x2-\x1,\y2-\y1)} in node{};

\calcLength(X1,Xtheta){mylen}

\path(O) -- (Xtheta) -- ([turn]\angleRefl:\mylen pt)node (Xrefl){};

\begin{scope}[very thick,decoration={
    markings,
    mark=at position 0.5 with {\arrow[black]{>}}}
    ] 
    \draw [very thick, postaction={decorate}] (Xtheta) --  node[left]{{\color{black}$\bar{U}_{t_i}-2(\bar{U}_{t_i}\cdot n_\mathcal{D})n_\mathcal{D}$}}(Xrefl);
\end{scope} 

\draw[very thick, shorten >=-0.1cm](Xtheta)--(Xrefl); %hack to extend the line
\node at (Xrefl)[cross=4pt, very thick]{};
\node at (Xrefl)[above=0.2cm]{${\bar{X}_{t_{i+1}}}$};

\draw[dotted, thick,-latex](O)--(Xtheta)-- ([turn]360:2cm)node[left]{$n_{\mathcal{D}}$}; 

\draw[-latex, very thin, loosely dashed, shorten >=0.3cm](6,2)node[above]{$\bar{X}_{\theta_i}$}to[out=-120,in=10] (Xtheta);

\end{tikzpicture}
\caption{Multidimensional scheme in the configuration space}
\label{fig:multi_d_scheme}
\end{figure}

Concerning the velocity component, a similar algorithm to Eqn. \eqref{eq:Velocity_discretization} is considered. The velocity process diffuses up to the collision time $\theta_i$, previously defined. At that time, the vector is reflected as such: 

\[
\bar{U}_{\theta_i} \leftarrow \bar{U}_{\theta_i^-} - 2\left(\bar{U}_{\theta_i^-}\cdot n_{\mathcal{D}}(\bar{X}_{\theta_i}) \right)n_{\mathcal{D}}(\bar{X}_{\theta_i})
\]

\subsection{Absorption scheme}\label{ssec:Absorption_scheme}

Besides the specular reflection problem, one can also consider the situation when two particles agglomerate after collision. Therefore the relative position and velocity both go to zero. This can be modeled by a process that has an absorption border at $\partial \mathcal{D} = \{|x| = \delta\}$. \red{The corresponding stochastic model is:}

	\begin{equation}\label{eq:Langevin_absorption_1D}
	\begin{dcases}
		& X_t = x+ \int_0^t U_s\,ds\quad \text{ if } t< \tau_{\text{abs}}\\	
		& U_t = u + \int_0^t{\color{black} b(s, X_s, U_s)}\,ds + \sigma W_t\quad \text{ if } t< \tau_{\text{abs}}\\
	\end{dcases}
	\end{equation}

\noindent where $\tau_{\text{abs}} \coloneqq \inf\{t > 0 ~;~ |X_t| = \delta\}$ is the absorption time. 

\red{The scheme is similar to the reflection result, except that it stops at the first hitting time $\tau_{\text{abs}} $.} If up to time $t_{i}$, the border hasn't been hit, then at time $t_i$ one can calculate $\theta_i$ defined in Eqn. \eqref{eq:Hitting_time}. If $t_i < \theta_i < t_{i+1}$, then the absorption border is hit and we return $\bar{X}_{\theta_i}= 0$ and $\bar{U}_{\theta_i^-}$ defined in Eqn. \eqref{eq:Velocity_discretization}.

Some  numerical results about the order of convergence are presented in the next section.

\section{Numerical experiments}\label{sec:Numeric}

Since the bias \eqref{eq:Error} cannot be calculated in a closed form, Monte Carlo simulation are carried out. As such, a statistical error is produced which dominates the bias for small enough time increments. Also the expectation of the exact process $\Ee f(X_T, U_T)$ is replaced with a reference result presented in subsection~\ref{ssec:Reference_result}. 
 
\begin{figure}[H]
\centering
\includegraphics[width=0.5\textwidth]{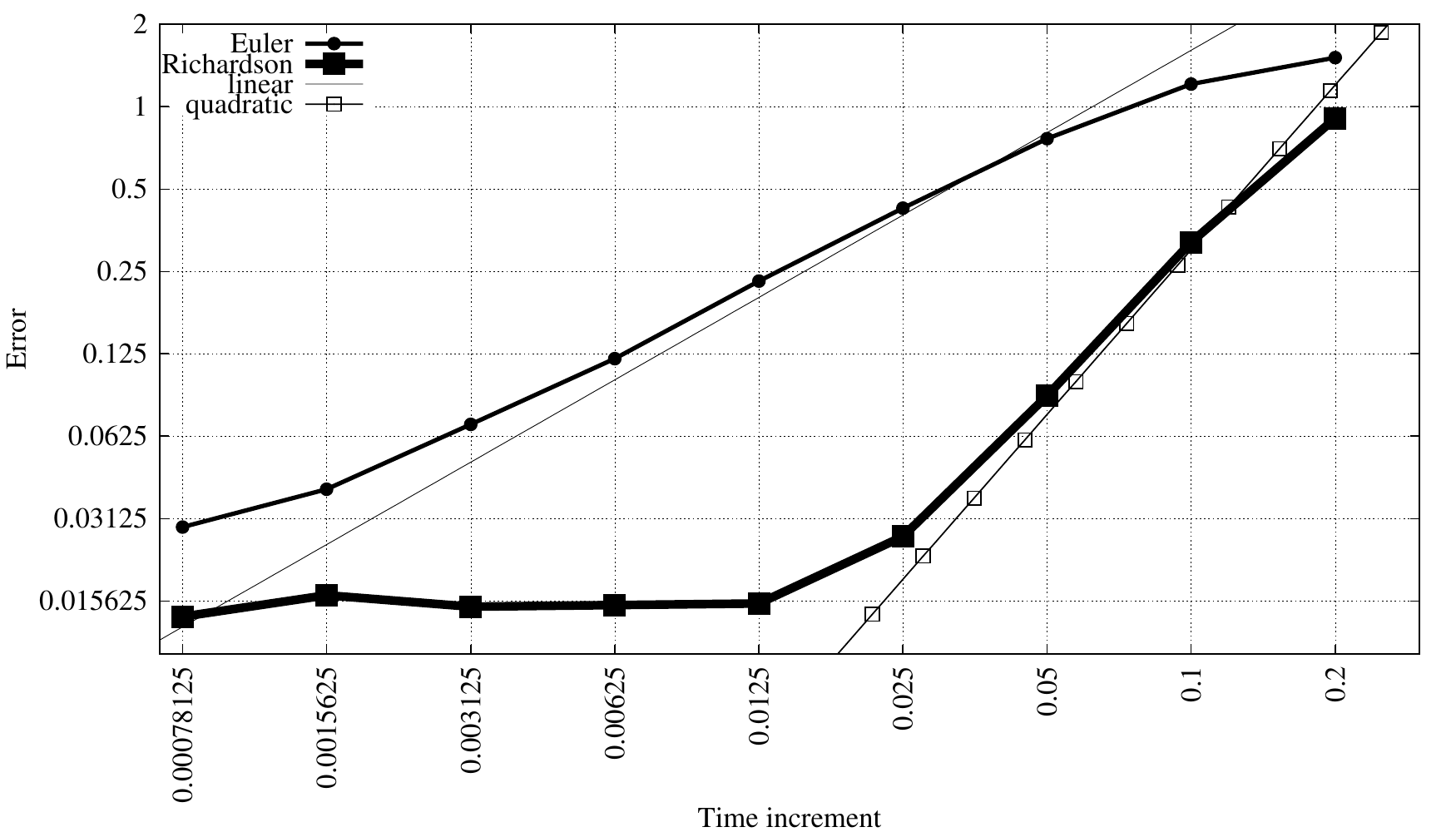}
\caption{Specular reflection results}
\label{fig:Specular_results}
\end{figure}

The Monte Carlo method simulates the scheme \eqref{eq:Position_discretization} - \eqref{eq:Velocity_discretization} on different partitions of $[0, T]$ with decreasing time increments~$\Delta t$.  
Figure \ref{fig:Specular_results} presents the results obtained, \red{in a log log plot}. For each simulation, $\Delta t$ is divided by two and the curve with round points is the total error:

\begin{align}\label{eq:Bias_Variance}
&\Ee \left(\Ee f(X_T, U_T) - \frac{1}{N}\sum_{n=1}^N f(\bar{X}^n_T, \bar{U}^n_T) \right)^2 = \text{Bias}[f](\Delta t)^2 + \notag\\
& \qquad + \Ee \left( \Ee f(\bar{X}_T, \bar{U}_T) - \frac{1}{N}\sum_{n=1}^N f(\bar{X}^n_T, \bar{U}^n_T) \right)^2
\end{align}
where $N$ is the number of trajectories, $(\bar{X}^n_T, \bar{U}^n_T)_{n \in \{1, \ldots, N\}}$ are independent realization of $(\bar{X}_T, \bar{U}_T)$.

Equation \eqref{eq:Bias_Variance} is called the bias-variance decomposition, the last term being the statistical error we commit.

 One can see that the total error approaches the simple line that represents a theoretical linear decrease of the error. When the time increment is reduced, the rate of decrease of the error degrades from the linear case as the statistical error begins to dominate the bias. At the start of the curve the slope does not correspond to a linear decrease though, so further analysis is needed. 

During the simulation, a trajectory hit the specular reflection border on average 1.14 times. This collision rate is stable as $\Delta t$ varies. 

\red{Since the time increment is halved for each simulation, is it possible to also perform, on the results, a Richardson extrapolation as in Ref. \cite{GrTa13}}. The error for the Richardson extrapolation error is the curve with full squares in Fig. \ref{fig:Specular_results}. 

Supposing that \eqref{eq:Error_estimate} holds, then theoretically, the Richardson extrapolation error should decrease with a quadratic speed, plotted by the line with empty square points. One can see that after a certain point, the curve flattens, thus at that point the statistical error becomes larger than the bias and decreasing the time step does not produce any improvement on the total error.

\subsection{Absorbed results}

The same type of simulation was calculated for the absorbed scheme from subsection~\ref{ssec:Absorption_scheme}. The reference result was calculated using a finer Monte Carlo method. Again one can notice that the error  decreases in a linear manner and the Richardson extrapolation, in the full square line, decreases quadratically.  

\begin{figure}[H]
\centering
\includegraphics[width=0.5\textwidth]{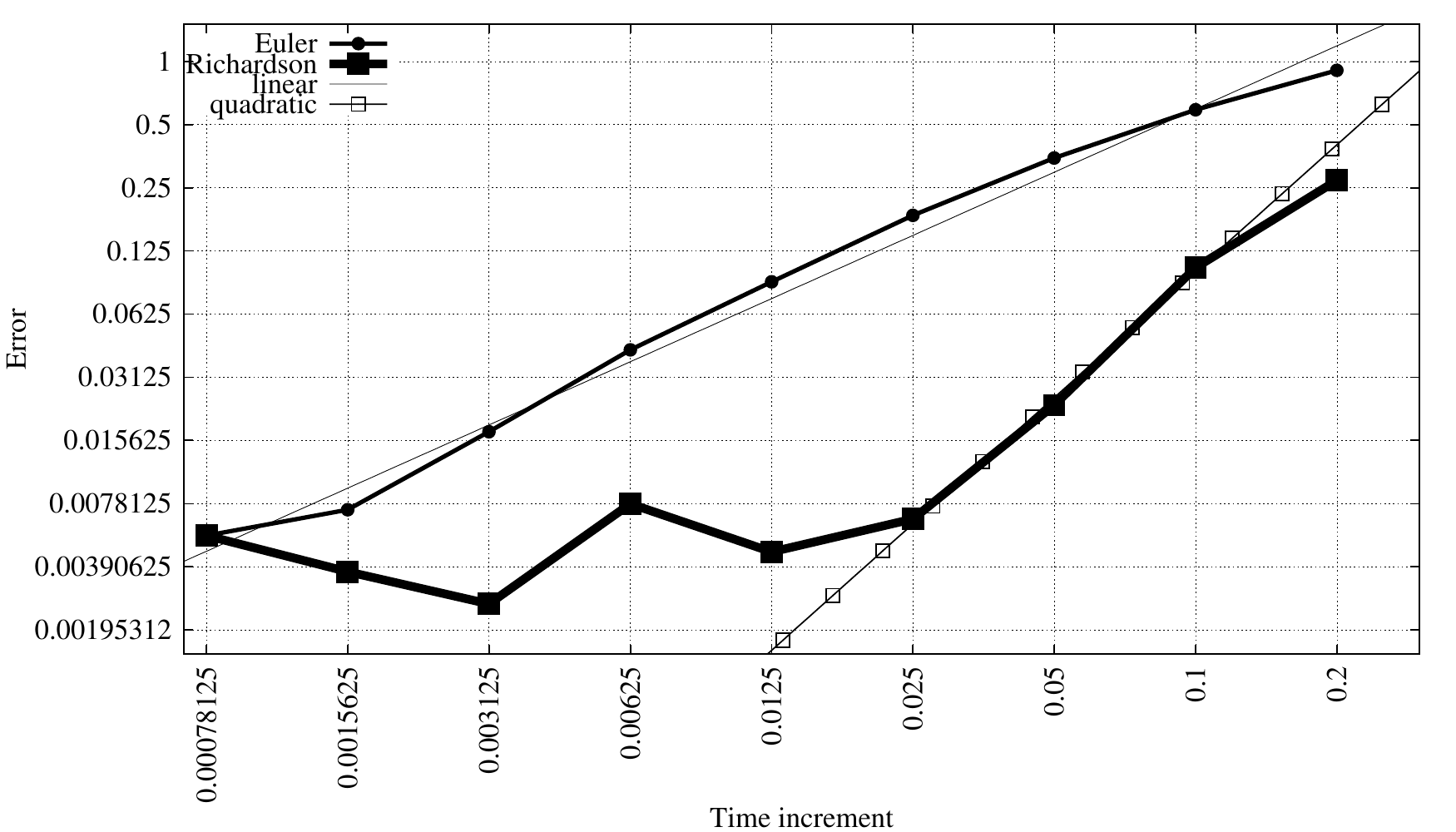}
\caption{Absorption results}
\label{fig:Abs_results}
\end{figure}

An average of 30\% of trajectories hit the absorption barrier placed at $x=0$.

\subsection{Details about the simulation}\label{ssec:Reference_result}

The drift was chosen to be:

\[
b(t, x, u) = \cos(2\pi x) + \frac{1}{2}\cos(2\pi u)
\]
\noindent the diffusion coefficient was constant $\sigma\equiv 1$ and the function $f$:

\[
f(x, u) = \left(10 - u \right)^2 \cdot (1-x)
\]

The initial condition for the specular reflection is $(x_0, u_0) = (0.5, -1.5)$ and for the absorption $(x_0, u_0) = (0.5, 1)$. 
The number of trajectories that were generated for the results in the previous sections is $10^9$ . The width of the 95\% confidence interval for both the specular reflection simulations and absorption simulations is of order $3 \cdot 10^{-3}$. The simulation is done for final time $T=3.2$. A periodic border is introduced at $x=1$ to not let the particles diffuse too far. So the position process was defined on $[0, 1]$.

The reference result for the specular reflection simulation was calculated by solving the PDE \eqref{eq:Main_PDE} using an implicit solver scheme. The time discretization step was taken to be of order~$10^{-4}$ while the position and velocity discretization steps are both of order~$5\cdot10^{-4}$. Because of numerical considerations, the velocity component was cut off for large values $u = \pm 10$ producing artificial boundary conditions for the PDE. The same velocity boundary conditions were imposed on the Monte Carlo method, therefore the velocity process was defined on $[-10~,10]$.

The reference result for the absorption scheme was obtained by Monte Carlo with the smallest time increment that we used: $\Delta t = 2.5 \cdot 2^{-10}$.

The value of the reference result for the specular case is: $49.8609$ and for the absorption case is: $64.5406$. 
\section{Conclusion}\label{sec:Conclusion}

By changing the reference frame, colloidal particles that experience perfectly elastic collisions can be modeled using Langevin models with specular reflection conditions. The article presents a discretization scheme and conjectures that the bias can be expanded as in \eqref{eq:Error_estimate} in the time increment $\Delta t$. 

Numerically, these conjectures are confirmed for the specular reflection scheme and also for the absorption scheme which model perfect agglomeration. 

The algorithm applies for a wide range of velocity drifts $b$ and can be extended to many particles. It has already been shown to be robust in a turbulent environment in the context of fluid particles. 

More analysis may allow to estimate the number of possible of collisions in order to adapt the time increment $\Delta t$. This would help with the larger goal of understanding how the position component in Langevin process with specular reflection converges towards a reflected Brownian motion. 

\section*{Appendix}\label{sec:Appendix}

If we assume a homogeneous fluid flow then we can integrate equation \eqref{eq:Langevin_model} to obtain:

\begin{equation}\label{eq:Lagevin_Exact}
\begin{dcases}
x_t =&\!\!\!\!  \tau_p\left(1 - \exp\left(-\frac{t}{\tau_p} \right) \right)u_0 + \\
	&\!\!\!\! + \int_0^t \left(1- \exp\left(-\frac{t-s}{\tau_p} \right) \right) \mathcal{U}_s \,ds\\
	&\!\!\!\! + B\tau_p W_t - B \tau_p \int_0^t \exp\left(-\frac{t-s}{\tau_p} \right)\,dW_s\\
u_t =&\!\!\!\! u_0 \exp\left(-\frac{t}{\tau_p} \right) + \frac{1}{\tau_p} \int_0^t \mathcal{U}_s \exp\left(-\frac{t-s}{\tau_p} \right)\,ds + \\
	&\!\!\!\!+B \int_0^t \exp\left(-\frac{t-s}{\tau_p} \right)\,dW_s 
\end{dcases}
\end{equation}

\noindent We introduce the process 
\begin{equation}\label{eq:Limit_process}
Y_t = \tau_p u_0 + \int_0^t \left(1- \exp\left(-\frac{t-s}{\tau_p} \right) \right) \mathcal{U}_s \,ds  + B\tau_p W_t  
\end{equation}

\noindent Then for smooth enough fluid velocity flows we have that

\begin{equation}
\lim_{\tau_p \to 0} \Ee \left|x_t - Y_t \right| = 0
\end{equation}

In particular, for any $\varepsilon  < t$:

\begin{equation}\label{eq:Exponential_limit}
\lim_{\tau_p \to 0} \exp\left(\frac{\varepsilon}{\tau_p} \right)\left|\Ee x_t - \Ee Y_t \right| = 0
\end{equation}

One can notice therefore that the process $(Y_t)_{t\geq 0}$ is a Brownian motion with drift. 
%\nocite{}
\bibliographystyle{plainnat}
%\bibliographystyle{icmf_style}
%\bibliography{Biblio_centrale}

\end{document}